\documentstyle[aps,prc,psfig,manuscript]{revtex}

\begin{document}
\bibliographystyle{h-physrev}

\title{Analysis of reaction dynamics at RHIC in a combined parton/hadron
       transport approach}

\author{S.A. Bass\footnote{Feodor Lynen Fellow of the Alexander v. Humboldt
        Foundation}}
\address{
        Department of Physics\\
        Duke University\\
        Durham, NC, 27708-0305, USA
        }

\author{M.~Hofmann, M.~Bleicher, L.~Bravina\footnote{Alexander v. Humboldt
	Fellow}\footnote{on leave of absence from the Institute of Nuclear
	Physics, Moscow State University, Moscow, Russia}, E.~Zabrodin$^\ddagger$, 
	H.~St\"ocker and 
	W.~Greiner}
\address{
        Institut f\"ur Theoretische Physik \\
        Johann Wolfgang Goethe Universit\"at\\
        Robert Mayer Str. 8-10\\
        D-60054 Frankfurt am Main, Germany}

\maketitle

\begin{abstract}
We introduce a transport approach which combines partonic
and hadronic degrees of freedom on an equal footing and discuss 
the resulting reaction dynamics. The initial parton dynamics is
modeled in the framework of the parton cascade model, hadronization
is performed via a cluster hadronization model and configuration space
coalescence, and the hadronic phase is described by a microscopic
hadronic transport approach. The resulting reaction dynamics indicates
a strong influence of hadronic rescattering on the space-time pattern
of hadronic freeze-out and on the shape of transverse mass spectra.
Freeze-out times and transverse radii increase by
factors of 2 -- 3 depending on the hadron species.
\end{abstract}

\pagebreak

Ultra-relativistic heavy ion collisions offer the unique opportunity
to probe highly excited dense nuclear matter under controlled laboratory
conditions and probably recreate a Quark Gluon Plasma (QGP), 
the highly excited state of 
primordial matter which is believed to have existed
shortly after the creation of the universe in 
the Big Bang (for recent reviews on the QGP, 
we refer to \cite{harris96a,bass98a}). 

While there is currently a strong debate whether
deconfined matter may have been produced at the CERN/SPS, it is widely expected
that in collisions of heavy nuclei at RHIC a QGP will be formed.
However, the deconfined quanta of a QGP are not directly
observable because of the fundamental confining property of the
physical QCD vacuum. What is observable are hadronic and
leptonic residues of the transient QGP state. 

Transport theory has been among the most successful approaches applied
to the theoretical investigation of relativistic heavy ion
collisions. Microscopic transport models attempt to describe the 
full time-evolution from the initial state of the heavy ion 
reaction (i.e. the two colliding nuclei) up to the freeze-out 
of all initial and produced particles after the reaction. The consequence
for the RHIC energy domain is that both, partonic and hadronic, degrees
of freedom have to be treated explicitly.

In this letter we introduce a transport approach which combines partonic
and hadronic degrees of freedom on an equal footing and discuss 
the resulting reaction dynamics. 
In terms of physics concepts the time evolution of the heavy ion collision
can be subdivided into three reaction stages: partonic evolution, hadronization
and hadronic evolution/freeze-out.

The initial partonic stage of the
reaction is modeled according to the Parton Cascade Model (PCM)
\cite{geiger92a,geiger95a}:
The nucleons of the colliding nuclei are resolved
into their parton substructure according to the measured nucleon
structure functions and yield the initial parton distributions.
Parton interactions as described by
perturbative QCD are used to model the evolution
of the ensemble of partons during the high energy density phase 
of the collision. 

To model hadronization, an effective QCD field theory
\cite{ellis95a} which fully describes the parton dynamics including
the formation of {\em preconfined} clusters is introduced.
Color singlet states are then formed via a configuration space coalescence
and a cluster ansatz \cite{geiger93a,webber94a}. 

For the hadronic phase of the collision 
a microscopic transport approach is employed 
\cite{peilert88a,aichelin91a,sorge89a,uqmdref1}, 
containing string- and hadronic degrees of freedom. 
All hadronic states can be produced in string decays, t-channel excitations,
s-channel collisions or resonance decays. Tabulated or parameterized experimental 
cross sections are used when available. Resonance absorption and scattering 
is handled via the principle of detailed balance. 

Note that our approach does not include any parton-hadron interactions: 
partons first hadronize when they cease to interact on a partonic level.
Therefore the newly formed hadronic states do not exert any pressure on
the partons and hadronization occurs as a unidirectional non-equilibrium process.

The technical realization of our approach uses
the framework of VNI \cite{geiger97a}\footnote{the current version of VNI, 
4.12, contains an erroneous rescaling of partons in coordinate space 
to the light-cone located at the origin of the computational 
coordinate system instead of that defined by the respective production vertex
of the parton. 
This erroneous rescaling has been removed for our calculations.}, 
both for the parton cascade model, 
as well as for the hadronization scheme.
The Ultra-relativistic Quantum Molecular Dynamics (UrQMD) model 
\cite{uqmdref1,uqmdref2} is employed  for the hadronic transport, decays and
rescattering.
The UrQMD model has been extensively tested in the SIS, AGS and SPS
energy domain \cite{uqmdref1}, it provides a robust description of hadronic
heavy-ion physics phenomenology.

Let us now turn to the reaction dynamics of central (impact parameter 
$b \le 1$~fm) Au+Au collisions at RHIC energies ($\sqrt{s}=200$~GeV 
per incident colliding nucleon-pair):
The initial reaction stage in VNI is dominated by the formation
of a dense partonic state.
Rapid thermalization is caused in VNI by radiative energy degradation and 
spatial separation of partons with widely different rapidities 
due to free streaming --
at RHIC, thermalization is predicted on a proper time scale
of approximately 1 fm/c \cite{eskola94a}.
The thermalized QGP is initially gluon rich and depleted of quarks due
to the larger cross section and higher branching ratios for gluons
\cite{shuryak92a}. 
The expansion then of course triggers the onset of hadronization,
leading to a phase  in which partonic and hadronic degrees of freedom
\cite{geiger98a} coexist.

\begin{figure}
\centerline{\psfig{figure=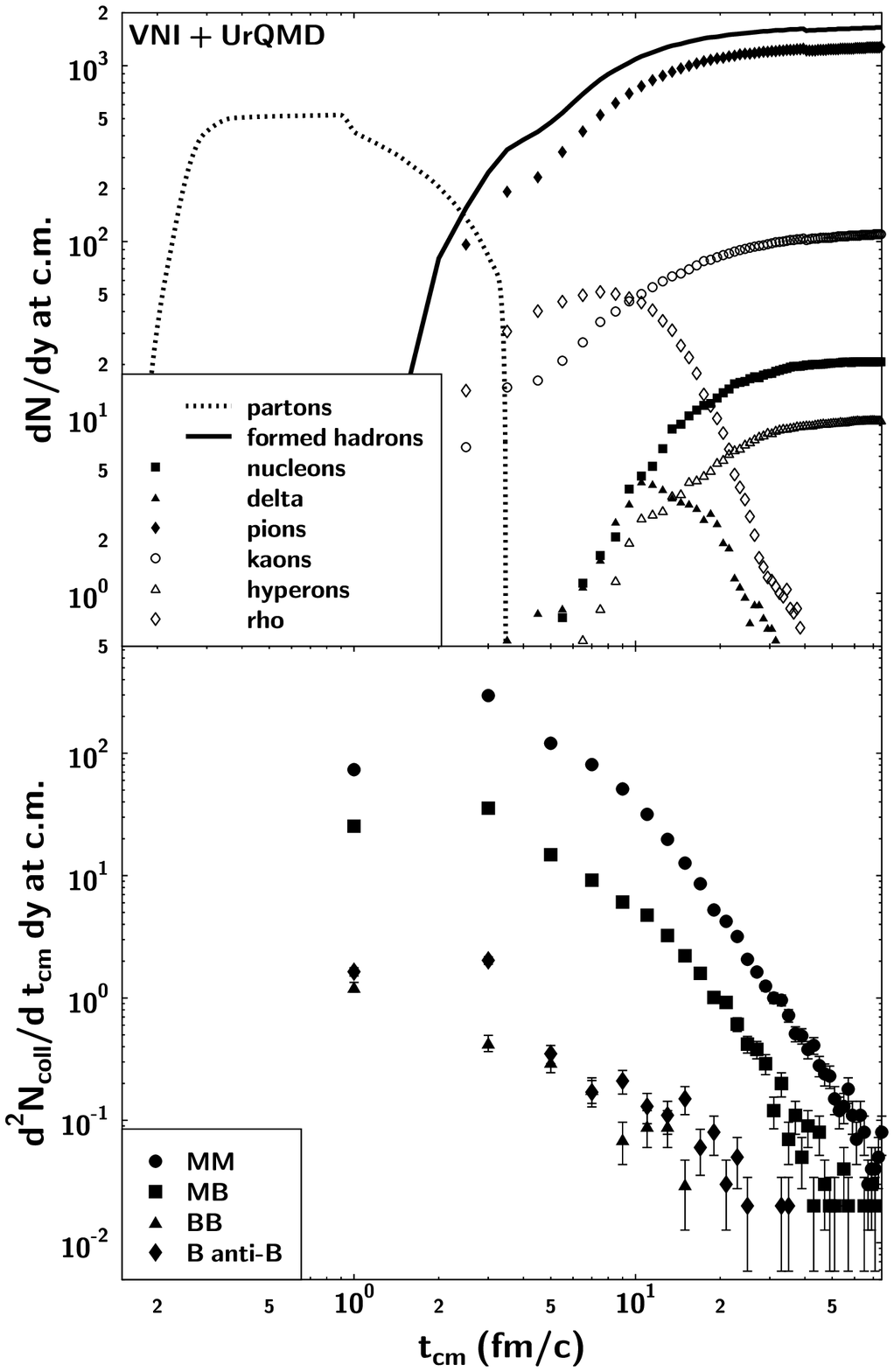,width=4.5in}}
\caption{\label{tevol} Top: time evolution of the parton and on-shell hadron
rapidity densities at c.m. for central ($b\le 1$~fm) Au+Au collisions at
RHIC. There exists a considerable overlap
between the partonic and hadronic phases of the reaction. Hadronic resonances
are formed and remain populated up to $\approx$~20~fm/c indicating a large
amount of hadronic interaction. Bottom: Rates for hadron-hadron collisions per
rapidity at c.m.. Meson-meson and to a lesser extent meson-baryon
interactions dominate the dynamics of the hadronic phase.}
\end{figure}

Starting point of our analysis is the time evolution (in c.m. time, 
$t_{c.m.}$)
of the rapidity density dN/dy of partons (i.e. quarks and gluons) 
and on-shell hadron
multiplicities at $|y_{c.m.}|\le 0.5$, depicted in 
the upper frame of figure~\ref{tevol}.
Note that there are no distinctly separate time scales for the three
reactions stages discussed earlier in this article: 
hadronic and partonic phases may evolve in
parallel and both, parton-parton as well as hadron-hadron interactions 
occur in the same space-time volume.
The overlap between the partonic and hadronic stages of the reaction 
stretches from $t_{c.m.} \approx 1$~fm/c up to $t_{c.m.} \approx 4$~fm/c
for the midrapidity region. Our analysis shows that this overlap occurs 
not only in time but also in coordinate space -- partonic and hadronic degrees
of freedom occupy the same space-time volume during this reaction phase.
Hadronic resonances like the $\Delta(1232)$ and the $\rho(770)$ (which
are the most abundantly produced baryonic and mesonic resonance states) 
are formed and remain populated up to 
$t_{c.m.} \approx 15 - 20$~fm/c, indicating a considerable
amount of hadronic rescattering. Hadron yields saturate
at time-scales $t_{c.m.} \approx 25$~fm/c. 
Since resonance decays have not been factored into this estimate
of the saturation time, this number should be viewed as an upper estimate for 
the time of chemical freeze-out.

Rates for hadron-hadron collisions
per unit rapidity at $y_{c.m.}$ 
are shown in the lower frame of figure~\ref{tevol},
i.e. all hadron-hadron collisions for hadrons with $|y_{c.m.}| \le 0.5$
were taken into account. 
Meson-meson and  meson-baryon
interactions dominate the dynamics of the hadronic phase. Due to their
larger cross sections baryon-antibaryon collisions occur more
frequently than baryon-baryon interactions. However, both are suppressed
as compared to meson-meson and meson-baryon interactions.
This is due to the large meson
multiplicity, which creates a  ``mesonic medium'' in which the
baryons propagate.

A comparison of calculations with and without hadronic rescattering shows that
e.g. the proton and antiproton multiplicities change by a factor of two due to 
hadronic rescattering, whereas the ratio of their yields remains roughly 
constant. Evidently chemical freeze-out of the system occurs well into the
hadronic phase and not at the ``phase-boundary''. The collision rates indicate
that interactions cease at $t_{c.m.} \approx 30-40$~fm/c 
at which point the system can be regarded as kinetically frozen out. Since
the saturation of the hadron yields occurs earlier, there is a clear separation
between chemical and kinetic freeze-out.

\begin{figure}
\centerline{\psfig{figure=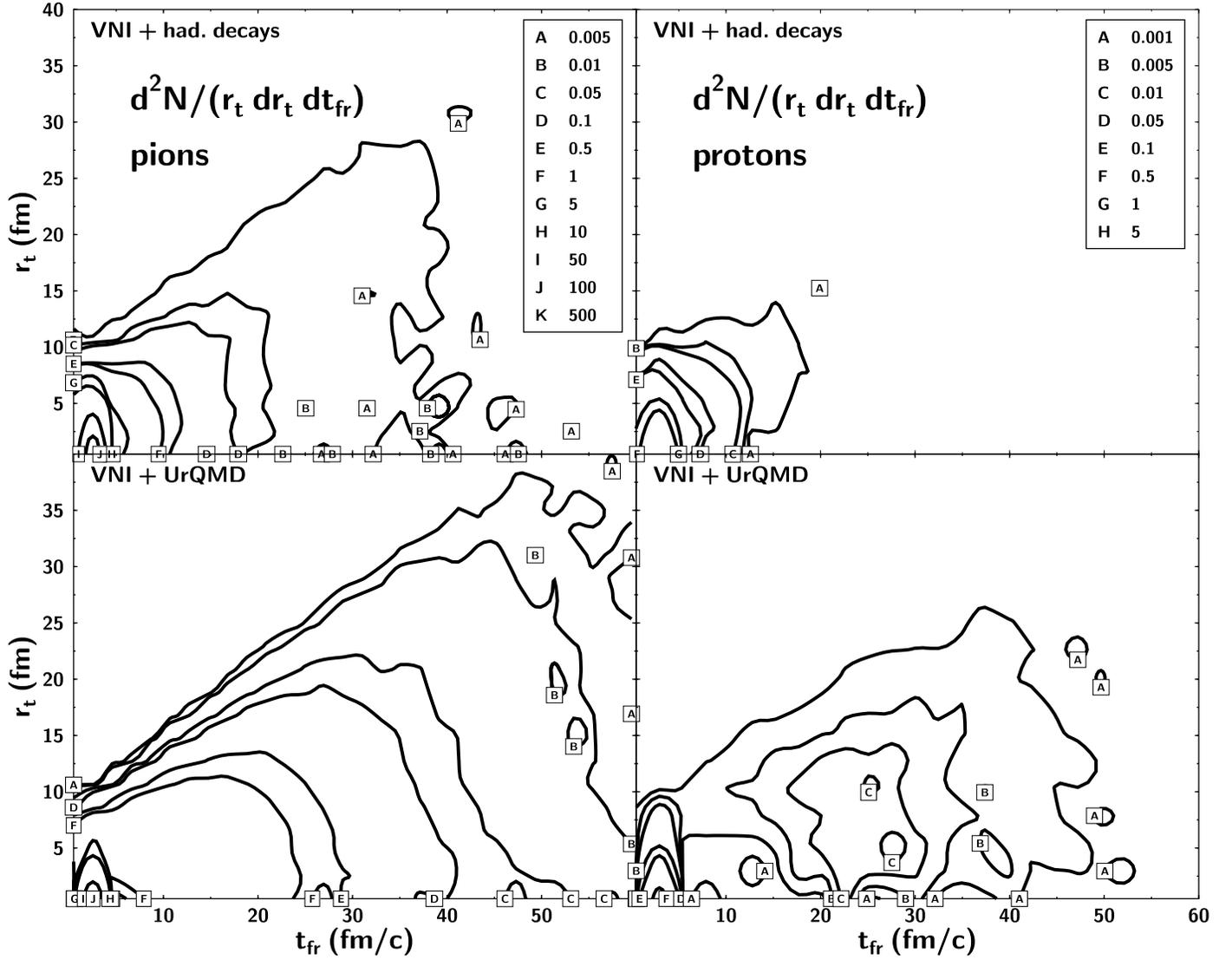,width=7.5in}}
\caption{\label{rttf} Freeze-out time and transverse radius distribution
d$^2N/(r_t$d$r_t$d$t_f)$ for pions (left column) and protons (right column)
around mid-rapidity ($-1 \le y_{c.m.}\le 1$).
The top row shows the result of parton cascading, hadronization and 
hadronic decays but without hadronic reinteraction whereas the lower row
shows the same calculation with full hadronic dynamics. The contour lines
have identical binning within each column but differ between the two columns.}
\end{figure}

Figure~\ref{rttf} shows freeze-out time $t_f$ and transverse radius $r_t$,
d$^2N/(r_t$d$r_t$d$t_f)$, for pions (left column) and protons (right column).
The top row shows the result of parton cascading, hadronization and 
hadronic decays, but without hadronic reinteraction, whereas the lower row
shows the same calculation with full hadronic dynamics. The contour lines
have identical binning within each column, but differ between the two columns.
Obviously hadronic interactions have an enormous effect on the freeze-out: 

The average freeze-out time for protons
increases tenfold from 3.2~fm/c without hadronic rescattering to 27.4~fm/c with
hadronic rescattering. The average transverse proton freeze-out
radius increases from 3.6~fm to 13~fm. 

For pions $t_f$ increases 
from 6.3~fm/c to 18.6~fm/c and $r_t$ from 5.6~fm to 9.6~fm!

Also, the overall shape of the distributions varies drastically: 
without hadronic
rescattering, the emission of pions and particularly protons is more strongly
restricted to the light-cone than with rescattering. This may give
rise to the speculation that HBT interferometry may indicate a surface-like
emission pattern in the case of negligible hadronic reinteraction and a
bulk-like emission pattern in the case of strong hadronic reinteraction. 
In the proton d$^2N/(r_t$d$r_t$d$t_f)$ distribution
hadronic rescattering
effects also manifest themselves through a second maximum 
at $t_{c.m.} \approx 25$~fm/c.
It should be noted, however, that even in the case of full hadronic rescattering,
approximately 15\% of protons and pions still freeze out directly after
hadronization without any hadronic (re)interaction or decay at all.

\begin{figure}
\centerline{\psfig{figure=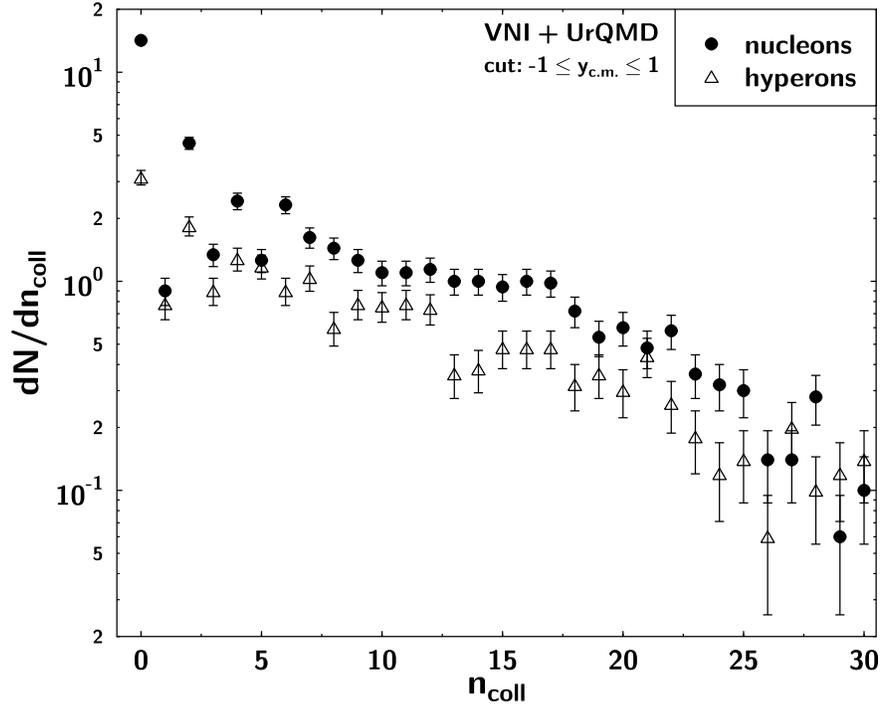,width=5in}}
\caption{\label{colldist} Collision number distribution for nucleons
and hyperons around midrapidity: 
approximately 15\% of the protons and 6\% of the hyperons
do not rescatter at all. The distribution exhibits a slow decrease from
$N_{coll}\approx 5$ to $N_{coll}\approx 17$ before falling  
roughly exponentially.}
\end{figure}

In contrast to the physics at the SIS and AGS, where 
(vector-)mesons propagate in a dense baryonic environment 
(allowing e.g. for a search for the onset of chiral symmetry restoration),
the situation at RHIC is reversed: here the baryons propagate
in a mesonic medium. Therefore we shall now focus on the dynamics of 
baryons in the intermediate and late reaction stages:
Figure~\ref{colldist} shows the collision number distribution for protons
and hyperons around midrapidity. 
The collision number distribution shows a very slow decrease from
$N_{coll}\approx 5$ to $N_{coll}\approx 17$, before falling off  
roughly exponentially.
Approximately 15\% of the protons and 6\% of the hyperons
do not rescatter at all. However, on average protons rescatter 14 and
hyperons 15 times. Baryonic rescattering is dominated (to 99\%) 
by meson-baryon interactions -- roughly 1/3 of those are elastic scatterings,
1/3 lead to the excitation of a discrete baryon resonance state with
a resonance-mass below 2.2~GeV and 1/3 lead to the excitation of a
state in the high-mass resonance continuum with masses above 2.2~GeV.
The rather small fraction of elastic scatterings serves as an indication
that the system does not fully thermalize in the course of the reaction.

\begin{figure}
\centerline{\psfig{figure=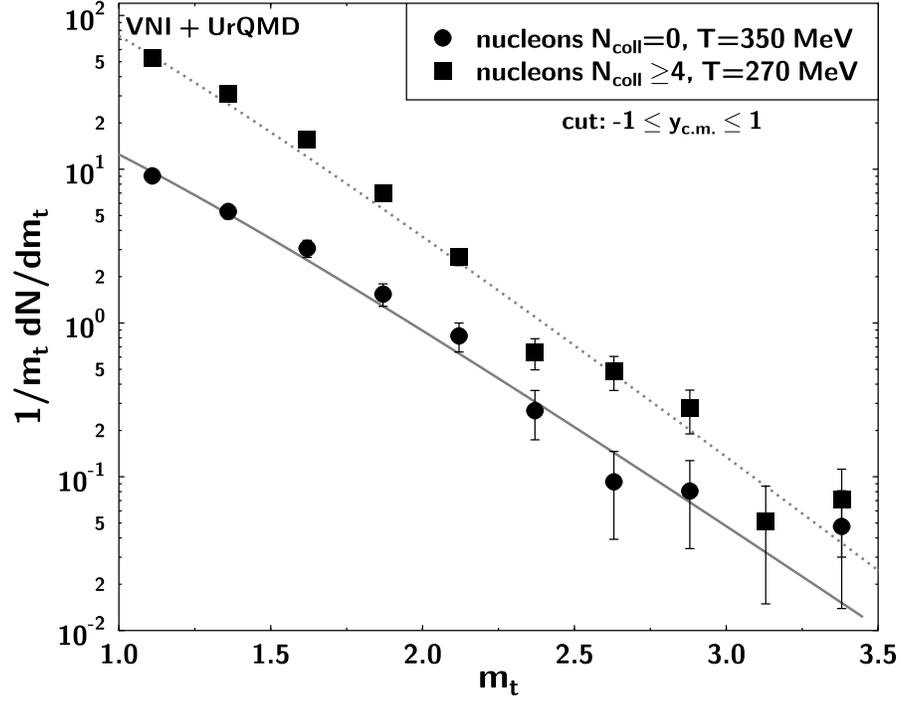,width=5in}}
\caption{\label{slope} d$N/m_t$d$m_t$ spectra for protons around midrapidity.
The circles show protons which have not rescattered after hadronization
whereas the squares refer to protons which have rescattered at least
4 times. The respective slope decreases by roughly 25\% indicating
a cooling of the protons via hadronic rescattering.}
\end{figure}

This massive hadronic rescattering does {\em not only} lead to changes in the
freeze-out space-time distribution,  but it also influences
the momentum distributions of the baryons: Figure~\ref{slope} shows the
d$N/m_t$d$m_t$-spectra for nucleons around midrapidity.
The circles denote nucleons which have not rescattered after hadronization,
whereas the squares refer to nucleons which have rescattered at least
4 times. The respective $m_t$ slope decrease by roughly 25\%
from $T_{N_c=0}=350$~MeV to $T_{N_c\ge4}=270$~MeV, thus
indicating a softening of the protons in the mesonic medium via rescattering.
This softening is quite surprising since other 
microscopic calculations at CERN/SPS and even RHIC energies in
the framework
UrQMD \cite{bleicher98a} and hybrid macro/microscopical \cite{dumitru99a}
models find massive rescattering leading to an
additional build-up of radial flow and a respective increase in the $m_t$ 
slope.

The effects of hadronic reinteraction on the predictions of 
perturbative QCD based transport descriptions \cite{geiger97a,wang91a} 
have also been investigated in \cite{geiger98a,bertsch88a,nara98a}.
In \cite{geiger98a,nara98a} the analysis was 
focused strongly on the influence 
of hadronic rescattering on inclusive hadronic spectra, whereas in 
\cite{bertsch88a} the partonic phase was not implemented in a 
fully microscopic fashion.
In contrast to \cite{geiger98a,bertsch88a,nara98a}, here we have
carried out an in-depth analysis of the underlying reaction dynamics in
the framework of a  combined microscopic parton/hadron transport 
approach.
Our results show that a radial flow analysis of hadrons and HBT correlation
measurements may be best suited to experimentally estimate the importance
of hadronic reinteractions at RHIC.
A detailed analysis of these experimentally accessible phenomena
is in preparation \cite{bass98vni2}.

In summary, 
we have introduced a novel transport approach combining partonic
and hadronic degrees of freedom and discussed 
the resulting reaction dynamics. 
At midrapidity partonic and hadronic degrees of freedom coexist between
$t_{c.m.} \approx 1$~fm/c and $t_{c.m.} \approx 5$~fm/c
in which numerous 
parton-parton and hadron-hadron interactions
take place. There exists no spatial separation between the partonic and 
hadronic interaction regions. 
The space-time pattern of particle freeze-out is strongly modified 
by hadronic rescattering. Freeze-out times and transverse radii increase by
factors of 2 -- 3 depending on the hadron species.
Massive rescattering of baryons in a highly excited meson gas is observed,
even though roughly 10\% of the baryons do not rescatter at all and are
directly emitted after hadronization. The rescattering also leads to a 
considerable softening in the baryon spectra.

This work is dedicated to the memory of Klaus Kinder-Geiger. 
It was made possible only through his active support 
and he would have surely been among
the coauthors if he was still alive.
S.A.B. acknowledges many helpful and inspiring discussions 
with Berndt M\"uller.
This work has been supported in part by the Alexander v. Humboldt Foundation
and in part by DOE grant DE-FG02-96ER40945, DFG, BMBF and the GSI 
Graduiertenkolleg ``Theoretische und Experimentelle Schwerionenphysik''.


\end{document}